\documentclass[a4paper,twoside,10pt]{article}
\usepackage{mathrsfs}
\usepackage[dvipdfm]{graphicx, color}
\input{jgrg19.sty}
\begin{document}

\pagestyle{fancy}
\fancyhead{}
  \fancyhead[RO,LE]{\thepage}
  \fancyhead[LO]{K. Saito}                  
  \fancyhead[RE]{CMB anisotropy in LTB cosmology}    
\rfoot{}
\cfoot{}
\lfoot{}
\label{O57}    

\def\Eq#1{\begin{equation} #1 \end{equation}}
\def\Eqr#1{\begin{eqnarray} #1 \end{eqnarray}}
\def\dfrac#1#2{\displaystyle\frac{#1}{#2}}
\def\lineq{\hspace{0.35em}\raisebox{0.4ex}{$<$}\hspace{-0.75em}\raisebox{-.7ex}{$\sim$}\hspace{0.3em}}
\def\gineq{\hspace{0.35em}\raisebox{0.4ex}{$>$}\hspace{-0.75em}\raisebox{-.7ex}{$\sim$}\hspace{0.3em}}
\def\p#1{\partial_{#1}}
\def\Ft{\p t F_0}
\def\Fr{\p r F_0}
\def\Fo{\p\omega F_0}
\def\Fm{\p\mu F_0}
\def\Fy{\omega\Fo}
\def\Fyy{(\omega\p\omega)^2 F_0}
\def\Fyr{\omega\p\omega\Fr}
\def\Fym{\omega\p\omega\Fm}
\def\Frr{\partial^2_r F_0}
\def\Frm{\p r \Fm}
\def\Fmm{\partial^2_{\mu}F_0}
\def\ape{a_{\perp}}
\def\apa{a_{\scriptscriptstyle /\! /}}
\def\Hpe{H_{\perp}}
\def\Hpa{H_{\scriptscriptstyle \! /\! /}}
\def\eP#1#2{e^{-\tilde P (#1, #2)}}
\def\eQ#1#2{e^{-\tilde Q (#1, #2)}}
\def\refsec#1{Sec. \ref{O57_sec: #1}}
\def\refeq#1{(\ref{O57_eq: #1})}
\def\reffig#1{Fig. \ref{O57_fig: #1}}
\def\ccite#1{\cite{O57_#1}}


\title{Analytic formulae for CMB anisotropy in LTB cosmology}

\author{Keiki Saito\footnote{Email address: saitok@post.kek.jp}$^{(a)}$,
			Akihiro Ishibashi$^{(b)}$
			and Hideo Kodama$^{(b)}$}

\address{$^{(a)}$Department of Particles and Nuclear Physics, \\
						The Graduate University for Advanced Studies (SOKENDAI), \\
						1-1 Oho, Tsukuba, Ibaraki 305-0801, Japan\\
				$^{(b)}$KEK Theory Center, Institute of Particle and Nuclear Studies, KEK, \\
							1-1 Oho, Tsukuba, Ibaraki, 305-0801, Japan}

\abstract{
The local void model has lately attracted considerable attention since 
it can explain the present apparent accelerated expansion of the 
universe without introducing dark energy.
However, in order to justify this model as an alternative cosmological
model to the standard $\Lambda$CDM model (FLRW universe plus dark energy),
one has to test the model by various observations, such as CMB 
temperature anisotropy, other than the distance-redshift relation of SNIa.
For this purpose, we derive some analytic formulae that can be used to 
rigorously compare consequences of this model with observations of CMB 
anisotropy and to place constraints on the position
of observers in the void model.
}

\section{Introduction}

In standard cosmology, we assume that our universe is isotropic and
homogeneous, and accordingly is described by the
Friedmann-Lema$\hat{\mbox{\i}}$tre-Robertson-Walker (FLRW) metric.
Recent observation of Cosmic Microwave Background (CMB) temperature
distribution on the celestial sphere shows that the spatial curvature is 
flat.
Furthermore, the distance-redshift relation of type Ia supernovae indicates
that the expansion of the present universe is accelerated.
Then, we are led to introduce, within the flat FLRW model,
``dark energy,'' which has negative pressure and behaves just like
a positive cosmological constant.
However, no satisfactory model that explains the origin of
dark energy has so far been proposed.


As an attempt to explain the SNIa distance-redshift relation without
invoking dark energy, Tomita proposed a ``local void model'' \ccite{Tomita1st}.
In this model, our universe is no longer assumed to be homogeneous,
having instead an underdense local void in the surrounding overdense 
universe.
The isotropic nature of cosmological observations is realized by assuming
the spherical symmetry and demanding that we live near the center of the 
void.
Furthermore, the model is supposed to contain only ordinary dust like 
cosmic
matter.  
Since such a spacetime can be described by
Lema$\hat{\mbox{\i}}$tre-Tolman-Bondi (LTB)
spacetime \ccite{L}-\ccite{B}, we also call this model the ``LTB cosmological model.''
Since the rate of expansion in the void region is larger than that
in the outer overdense region, it can explain the observed dimming of
SNIa luminosity.
In fact, many numerical analysis \ccite{Tomita1st2}-\ccite{GBHkSZ} have recently 
shown that this LTB model can accurately
reproduce the SNIa distance-redshift relation.

However, in order to verify the LTB model as a viable cosmological model,   
one has to test the LTB model by various observations---such as
CMB temperature anisotropy---other than the distance-redshift
relation\footnote{
Recently, some constraints on the LTB model from BAO and kSZ effects
have also been discussed, see e.g. \ccite{GBHkSZ}. Still,
the possibility of the LTB model is not completely excluded.
}.  
For this purpose, in this paper, we derive some analytic formulae that
can be used to rigorously compare consequences of the LTB model with
observations of CMB anisotropy. More precisely, we derive analytic formulae
for CMB temperature anisotropy for dipole and quadrupole momenta,
and then use the dipole formula to place the constraint on the distance
between an observer and the symmetry center of the LTB model.
We also check the consistency of our formulae with some numerical analysis
of the CMB anisotropy in the LTB model, previously made by Alnes
and Amarzguioui \ccite{AACMB}.


In \refsec{LTB}, we briefly summarize the LTB metric.
In \refsec{multipole}, we derive analytic formulae for CMB anisotropy in the LTB model.
In \refsec{constraint}, we obtain some constraints concerning the position of the observer.
\refsec{summary} is devoted to a summary.

\section{LTB spacetime}
\label{O57_sec: LTB}

	A spherically symmetric spacetime with only non-relativistic matter is described by the Lema$\hat{\mbox{\i}}$tre-Tolman-Bondi (LTB) metric \ccite{L}-\ccite{B}
\Eq{
	ds^2 = -dt^2 + \frac{\{R' (t, r)\}^2}{1-k(r)r^2}dr^2 + R^2 (t, r) d\Omega_2^2,
}
where $' \equiv \p r$, $k(r)$ is an arbitrary function of only $r$.
The Einstein equations reduce to
\Eqr{
	\left(\frac{\dot R}{R}\right)^2 &=& \frac{2GM(r)}{R^3} - \frac{k(r)r^2}{R^2}, \\
	4\pi\rho (t,r) &=& \frac{M' (r)}{R^2 R'},
}
where $\dot{} \equiv \p t$, $M(r)$ is an arbitrary function of only $r$, and $\rho(t, r)$ is the energy density of the non-relativistic matter.
The general solution for the Einstein equations in this model admits two arbitrary functions $k(r)$ and $M(r)$.
By appropriately choosing the profile of these functions, one can construct some models which can reproduce the distance-redshift relation of SNIa in this model.

\section{Analytic formulae for CMB anisotropy in LTB model}
\label{O57_sec: multipole}

	In this section, we derive analytic formulae for the CMB anisotropy in the LTB model.
First, we assumed that the universe was locally in thermal equilibrium (that is, the distribution function $F$ was Planck distribution $\Phi$) at the last scattering surface, and the direction of the CMB photon traveling is fixed.
In this case, $F$ can be written as $F = \Phi (\omega/T)$, where $\omega \equiv p^t$, and $T$ is the temperature.
Then, the CMB temperature anisotropy $\delta T/T$ is defined by
\Eq{
	\delta F = -\frac{\delta T}{T}\omega\p\omega F.
}
Second, supposing that an observer lives at a distance of $\delta x^i$ from the center of the void, it follows that
\Eqr{
	(\delta F)^{(1)} &=& \delta x^i (\p i F)_0, \\
	(\delta F)^{(2)} &=& \frac{1}{2}\delta x^i \delta x^j (\p i \p j F)_0,
}
where the subscript 0 means the value at the center ($r = 0$) at the present time ($t = t_0$).
From these, the CMB temperature anisotropy dipole $(\delta T/T)^{(1)}$ and quadrupole $(\delta T/T)^{(2)}$ are written as
\Eqr{
	\left(\frac{\delta T}{T}\right)^{(1)} &=& -\frac{\delta x^i (\p i F)_0}{\Fy}, \\
	\left(\frac{\delta T}{T}\right)^{(2)} &=& -\frac{1}{2}\frac{\delta x^i \delta x^j (\p i\p j F)_0}{\Fy}
														+\frac{1}{2}\left\{\left(\frac{\delta T}{T}\right)^{(1)}\right\}^2 \frac{\Fyy}{\Fy}.
}

	We assume that the distribution function $F(x, p)$ itself is spherically symmetric.
Then, $F$ can be written as $F(x, p) = F_0 (t, r, \omega, \mu)$, where $\mu \equiv R'p^r /(\sqrt{1 - kr^2}\omega)$.
This implies that $\p i F = (\p i r)\Fr + (\p i \omega)\Fo + (\p i \mu)\Fm$.
Then, we can derive analytic formulae for the CMB anisotropy dipole by solving the Boltzmann equation $\mathscr{L}[F_0] = \Ft + \dot r \Fr + \dot{\omega}\Fo + \dot{\mu}\Fm = 0$.
The result is 
\Eq{
	\left(\frac{\delta T}{T}\right)^{(1)} = \delta L n^j \Omega_j \left\{\frac{\sqrt{1 - k(r_i)r_i^2}}{R'_0}\eP{t_0}{t_i}\left(\frac{\Fr}{\Fy}\right)_i
																		+\int^{r_i}_0 dr \Hpa' \exp\left[\int^t_{t_0}dt_1 \Hpa(t_1)\right]\right\},
\label{O57_eq: dipole}
}
where $\delta L n^j$ is the position vector of the observer, $\Omega^j \equiv x^j /r$, $\tilde P(t_0, t_i) \equiv \int^{r_i}_0 dr R''/R'$, $\Hpa \equiv \dot R'/R'$, and the subscript $i$ denotes the value at the last scattering surface.
By a similar method, we also derive the CMB anisotropy quadrupole formula
\Eqr{
	\left(\frac{\delta T}{T}\right)^{(2)} &=& -\frac{\delta x^i \delta x^j}{2(\Fy)_i}
														\Biggl[(\delta_{ij} - \Omega_i \Omega_j )\left(\frac{\Fr}{r} - \mu\frac{\Fm}{r^2}\right)_0 + \Omega_i \Omega_j (\Frr)_0
														\nonumber\\
													&& \hspace{62pt} + \left\{\frac{\ape''}{\ape}\delta_{ij}
																						+\ape\left(\frac{R'}{\sqrt{1 - kr^2}} - \ape\right)''\frac{\Omega_i \Omega_j}{(R')^2}\right\}_0
																				(\Fy)_i \Biggr] \nonumber\\
													&& +\frac{1}{2}\left\{\left(\frac{\delta T}{T}\right)^{(1)}\right\}^2 \frac{\Fyy}{\Fy},
\label{O57_eq: quadrupole}
}
where $\ape \equiv R/r$.

\section{Constraint on LTB model}
\label{O57_sec: constraint}

	In this section, we derive some constraints concerning the position of the off-center observers in the LTB model from the CMB dipole formula \refeq{dipole}.
In general, the CMB temperature anisotropy is decomposed in terms of the spherical harmonics $Y_{lm}$ by
\Eq{
	\frac{\delta T}{T} = \sum_{l, m} a_{lm}Y_{lm},
}
where the amplitudes in the expression are recovered as
\Eq{
	a_{lm} = \int^{2\pi}_0 d\phi \int^{\pi}_0d\theta \sin\theta\frac{\delta T}{T}Y_{lm}.
}
We are interested in $a_{10}$ as the dipole moment.
We estimate the CMB dipole formula \refeq{dipole} numerically by using the profile considered in \ccite{AACMB} (\reffig{AA}),
\Eqr{
	M(r) &=& \frac{1}{2}H_{\perp}^{2}(t_{0}, r_{\rm out})r^{3}\left[\alpha_{0} - \Delta\alpha\left(\frac{1}{2}
																					- \frac{1}{2}\tanh{\frac{r - r_0}{2\Delta r}}\right)\right], \\
	k(r) &=& -H_{\perp}^{2}(t_{0}, r_{\rm out})\left[\beta_{0} - \Delta\beta\left(\frac{1}{2} - \frac{1}{2}\tanh{\frac{r - r_0}{2\Delta r}}\right)\right],
\label{O57_eq: profile1}
}
where
\Eqr{
	t_{s}(r) = 0, \ \ \Hpe(t_{0}, r_{\rm out}) = 51 \ {\rm km/s/Mpc}, \ \ \alpha_{0} = 1, \ \ \Delta\alpha = 0.90, \nonumber\\
	r_{0} = 1.34 \ {\rm Gpc}, \ \ \Delta r = 0.536 \ {\rm Gpc}, \ \ \beta_{0} = 1 - \alpha_{0} = 0, \ \ \Delta\beta = -\Delta\alpha = -0.90,
\label{O57_eq: profile2}
}
and $\Hpe \equiv \dot{\ape}/\ape$.
\begin{figure}
	\begin{tabular}{cc}
		\begin{minipage}{0.46\hsize}
			\begin{center}
				\includegraphics[width=0.77\hsize]{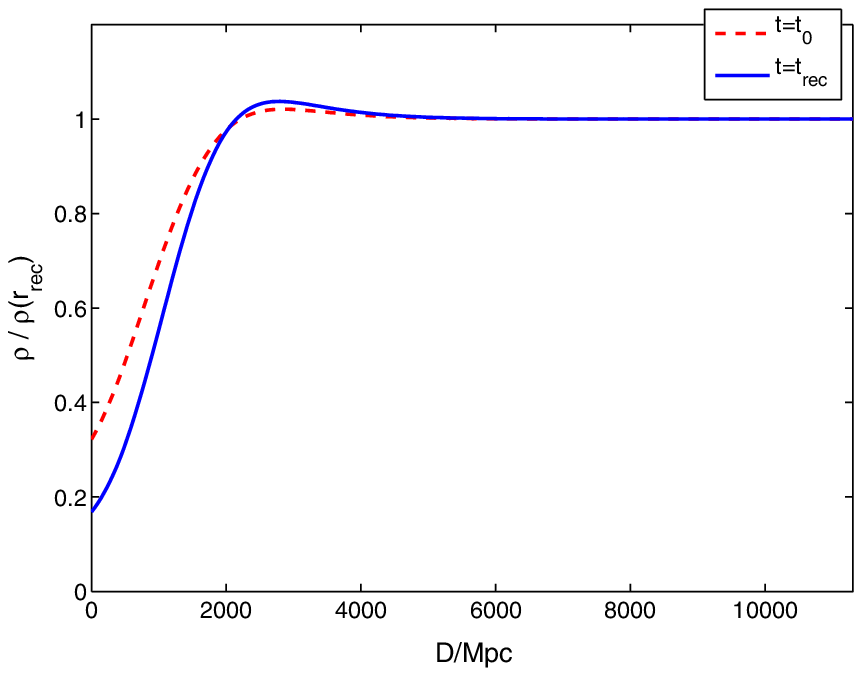}
			\end{center}
		\end{minipage}
		
		\begin{minipage}{0.48\hsize}
			\begin{center}
				\includegraphics[width=0.77\hsize]{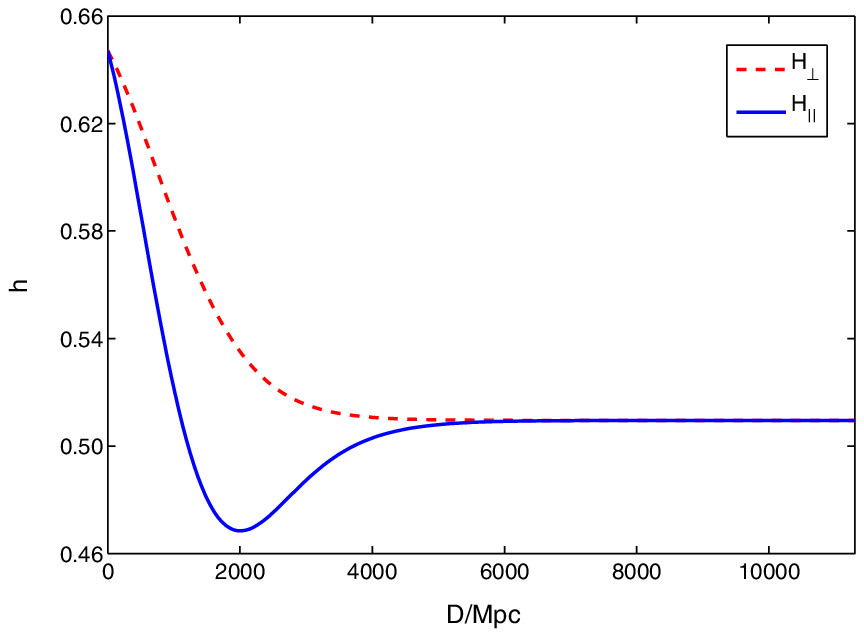}
			\end{center}
		\end{minipage}
	\end{tabular}
\caption{The profile considered in \ccite{AACMB}.
			The subscript $rec$ denotes the value at the recombination, and $\Hpa ({\rm or}\ \Hpe) = 100h \rm km/s/Mpc$. }
\label{O57_fig: AA}
\end{figure}
The induced $a_{10}$ is of order $10^{-3}$ or less observed by Cosmic Background Explorer (COBE) \ccite{COBE}, so we find that
\Eq{
	\delta L \lineq 15 \rm Mpc,
}
where $\delta L$ is the distance from the observer to the center of the void.
This is consist with the result of \ccite{AACMB}.

\section{Summary}
\label{O57_sec: summary}

	In the LTB model, we have derived the analytic formulae for the CMB anisotropy dipole \refeq{dipole} and quadrupole \refeq{quadrupole}, which can be used to rigorously compare consequences of this model with observations of the CMB anisotropy.
Moreover, we checked the consistency of our formulae with results of the numerical analysis in \ccite{AACMB}, and constrained the distance from an observer to the center of the void.
One of the advantages in obtaining analytic formulae is that we can identify physical origins of the CMB anisotropy in the LTB model.
For example, in the CMB dipole formula \refeq{dipole}, we can regard the first term as the initial condition at the last scattering surface, and the second term as the Integrated Sachs-Wolfe effect.

\section*{Acknowledgments}

	The authors would like to thank all participants of the workshop
$\Lambda$-LTB Cosmology (LLTB2009) held at KEK from 20 to 23  October
2009 for useful discussions.
We would also like to thank Hajime Goto for useful discussions.
This work is supported by the project
Shinryoiki of the SOKENDAI Hayama Center for Advanced Studies and the
MEXT Grant-in-Aid for Scientific Research on Innovative Areas (No.
21111006).


\end{document}